\documentstyle {article}

\def\rth{\frac{1}{2\sqrt{3}}}
\def\cha{\frac{1}{4\sqrt{3}}}
\def\rtw{\frac{1}{2\sqrt{2}}}
\def\rsi{\frac{1}{2\sqrt{6}}}
\def\rtwi{{2\sqrt{2}}}
\def\s{{\sqrt{2}}}
\def\rsii{{2\sqrt{6}}}

\def\t{\tilde}

\def\b{\bar}
\def\e{\epsilon}

\def\be{\begin{equation}}
\def\bea{\begin{eqnarray}}
\def\t{\tilde}
\def\dbar{\bar \partial}
\def\nn{\nonumber}

\def\ee{\end{equation}}
\def\eea{\end{eqnarray}}
\def\w{WZW }
\def\o{\over}
\begin{document}
\begin{titlepage}
\hfill
\vbox{
    \halign{#\hfil         \cr
           hep-th/9811009\cr
           %IPM/P-98/??\cr
           } % end of \halign
      }  % end of \vbox
\vspace*{3mm}
\begin{center}
%%%%%%%%%%%%%%%%%%%%  The title page   %%%%%%%%%%%%%%%%%%%%%%%%%%%%%%%%%%%%%
\hfill
\vskip 5 mm
{\large \bf  On the Exceptional Gauged WZW Theories}
\vskip .6in
{\bf Amir Masoud Ghezelbash} \footnote{e-mail address:
amasoud@theory.ipm.ac.ir}\\
\vskip .25in
{\em
Department of Physics, Alzahra University, Vanak,
Tehran 19834, Iran.}\\
{\em  and}\\
{\em
Institute for Studies in Theoretical Physics and Mathematics,\\
P.O. Box 19395-5531, Tehran, Iran.}\\
\vskip .5in
\end{center}
\begin{abstract}
 We consider two different versions of gauged \w theories with the
exceptional groups
and gauged with any of theirs null subgroups. By constructing suitable
automorphism, we establish the equivalence of these two theories. On the
other hand our automorphism, relates the two dual irreducible Riemannian
globally symmetric spaces with different
characters based on the corresponding exceptional Lie groups.
\vskip .5in
%PACS: 11.15.-q,02.20.+b
\end{abstract}
\end{titlepage}
\newpage
%%%%%%%%%%%%%%%%%%% The body of the paper %%%%%%%%%%%%%%%%%%%%%%%%%%%%%%%%%%%
\noindent{\bf{Introduction}}

In the last few years, duality transformation as an abelian or non-abelian
symmetry of conformal field theories has
been studied
extensively \cite{1,11,12}.
In \w models, the duality transformations are given by the automorphisms of the
group $G$ of the models. In particular,
it has been shown that
the target space of the gauged \w model with the group $G$ and vector gauged
abelian subgroup is dual to the corresponding target space of the axial gauged
\w model gauged by the same abelian subgroup.
Moreover the
duality transformation is
implemented by an automorphism of the gauged subgroup \cite{2,21,22,23}.
When the gauged subgroup is not
semisimple, and in particular when it is null,
the target space of vector gauged \w model reduces to a space with
less dimensions
\cite{3,31,32,33}.
In other words, it was shown that in the usual vector gauging of classical Lie
group $G$ \w model, by its maximal null subgroup $H$, the effective action of
the gauged model reduces to that of a Toda theory which the
number of Toda fields is less than Dim$G$-Dim$H$.
An explanation for the reduction of the degrees of freedom of the target space
of the vector gauged model is
related to the obvious dimensional reduction of the
corresponding chiral gauged model when the left and right gauge actions are
independent and in two different subgroups isomorphic to vector null gauged
subgroup.
In ref. \cite{4}, it was proved that the vector and chiral gauged theories
with any classical Lie groups and any null gauged subgroup are equivalent
to each other. The equivalence map was constructed for every classical
Lie groups and it was shown that these mapps were in fact involutive
automorphisms of corresponding Lie algebras. These involutive
automorphisms could
be used for real form construction of simple Lie algebras and also gave us
the pairs of dual Riemannian globally symmetric spaces \cite{5}.\newline
In this paper, we consider the vector and chiral gauged \w models which
based on the exceptional groups and gauged with any of theirs null subgroups
and prove that these models are equivalent to each other. We will find that
the equivalence map which relates the two models, is the involutive automorphism
of the algebra and in the context of Riemannian geometry relates the
two dual irreducible Riemannian globally
symmetric spaces based on the exceptional Lie group. In particular, for the
exceptional group $G_2$, we give the calculation in detail, and in the case of
other exceptional groups, because of complexity, we present the final results.
The work done in this paper completes the equivalence of vector and chiral
gauged \w models with any Lie groups.
\vskip 0.5 cm
\noindent{\bf{$G_2$ Gauged WZW Models}}

Let's recall the structure of gauged \w models based on the exceptional Lie group $G_2$. The vector gauged
action is given by \cite{6,61}
\bea S_V(g,{\bf {A}},{\bf {\bar A}}
)=S(g)+{k\over 2\pi}\,\int
\,d^2z\,
Tr(-\,{\bf {\bar A}} g^{-1} \partial g+{\bf {A}}\dbar {g} g^{-1}-
{\bf {A}}{\bf {\bar
A}}+g{\bf {\bar A}}g^{-1}{\bf {A}}\,), \eea $$
S(g)={k\over 4\pi}\,\int
\,d^2z\,
Tr(\,g^{-1}\partial gg^{-1} \bar \partial
g\,)-{k\over 12\pi}\,\int \,
Tr(\,g^{-1}dg\,)^3,
$$
which
is invariant under the gauge transformations
\be g \rightarrow h^{-1}g\,h,{\bf {A}}\rightarrow h^{-1}\,({\bf
{A}}-\partial)\,h,{\bf {\bar A}}\rightarrow h^{-1}\,({\bf {\bar A}}-\bar
\partial)\,h,
\ee where $ h=h(z,\bar z) $ is a group element in subgroup $ H $ of $G_2$ and
$ {\bf {A}}$ and $ {\bf {\bar A}} $ take their values in the algebra
${\cal L}(H) $ of subgroup $ H $.
On the other hand, the chiral \w action \cite{7}
\bea S_C(g,{\bf {A}},{\bf {\bar A}})=S(g)+{k\over 2\pi}\,\int
\,d^2z\,
Tr(-\,{\bf {\bar A}} g^{-1} \partial g+{\bf {A}}\dbar {g} g^{-1}+g{\bf {\bar
A}}g^{-1}{\bf {A}}\,), \eea
is invariant under the following transformations
\be  g \rightarrow h^{-1}g\,\bar h,{\bf {A}}\rightarrow
h^{-1}\,({\bf {A}}-\partial)\,h,{\bf {\bar A}}
\rightarrow \bar h^{-1}\,({\bf {\bar A}}-\bar \partial)\,\bar h, \ee
where $ h=h(z) $ belongs to the subgroup $ H_1 $, and $ \bar h=\bar h(\bar z) $
belongs to another subgroup $ H_2 $ of $ G_2 $. $ {\bf {A}}$ takes its value in
$\cal {L}$
$(H_1)$ and $ {\bf {\bar A}} $ in $ \cal {L}$ $(H_2)$.\newline
Now, we impose following transformations on the
$g$ field and the gauge fields of the chiral theory
\be \label{THETA}
g'=g\theta,\quad{\b A}'=\theta ^{-1}{\b A}\theta,\quad A'=A,
\ee
and demand that the $g'$ and ${\b A}'$ become the corresponding fields
in the vector theory.
\newline
After straight forward but lengthy calculations, one can find that
\be
\theta=\pmatrix{0&0&0&1&0&0&0\cr0&0&1&0&0&0&0\cr0&1&0&0&0&0&0\cr
1&0&0&0&0&0&0\cr0&0&0&0&0&0&-1\cr0&0&0&0&0&-1&0\cr0&0&0&0&-1&0&0
}.
\ee
It is obvious that $\theta ^2=1$. For proving the equivalence,
it is also necessary that the involutive
automorphism $\theta$ belongs to the Lie group $G_2$. According to
a well known theorem \cite{5},
an involutive automorphism of a compact semisimple Lie algebra ${\cal{A}}$
belongs to the adjoint group of ${\cal{A}}$ (inner automorphism)
if and only if the rank of the set of
fixed points of the automorphism is equal to the rank of the Lie algebra ${\cal{A}}$
. Let us take $E$ to be in the Lie algebra $g_2$, according to
\bea E&=&
\rsii (e_1 E_1+\tilde e_1 E_{-1}+e_3 E_3+\tilde e_3 E_{-3}+e_5 E_5+\tilde e_5
E_{-5}
+e_6 E_6+\tilde e_6 E_{-6})\nn \\&+&\rtwi (e_2 E_2+\tilde e_2 E_{-2}
+e_4 E_4+\tilde e_4 E_{-4})+4\sqrt 3h_1H_1+4h_2H_2,
\eea
where $E$'s, $H_1$ and $H_2$ are given in the Appendix. Then, the invariant set
of involutive automorphism $\theta$ is $f={{E+\theta E\theta}
\o 2}$, which with the redefinition of variables
\be
\epsilon _1={{e_1+\t e_1} \o 2},\epsilon _2={{e_2-\t e_2} \o 2},
\epsilon _3={{e_3+\t e_3} \o 2},\epsilon _4={{e_4+\t e_4} \o 2},
\epsilon _5={{e_5-\t e_5} \o 2},\epsilon _6={{e_6+\t e_6} \o 2},
\ee
can be written as
\be
f=\pmatrix{0&\e _1&\e _4&0&-\e _5&\sqrt 2 \e_3&\e _2\cr
\e _1&0&0&\e _4&-\e _6&\sqrt 2 \e_5&-\e _3\cr
\e _4&0&0&\e _1&\e _3&-\sqrt 2 \e_5&\e _6\cr
0&\e _4&\e _1&0&-\e _2&-\sqrt 2 \e_3&\e _5\cr
\e _5&-\e _6&\e _3&\e _2&0&\sqrt 2 \e_1&0\cr
\sqrt 2\e _3&-\sqrt 2\e _5&\sqrt 2\e _5&-\sqrt 2\e _3&\sqrt 2\e _1&0&
\sqrt 2\e _1\cr
-\e _2&-\e _3&\e _6&-\e _5&0&\sqrt 2 \e_1&0
}.
\ee
It can easily be shown that $f$ produces an algebra and its rank is two.
More precisely, the algebra $f$ is isomorphic to $su(2)\otimes su(2)$.
Hence, $\theta$ belongs to $G_2$ Lie group and establishes the equivalence of
vector
and chiral theories with $G_2$ group. It is also interesting to note the relation
of our
constructed automorphism $\theta$ and the dual Riemannian globally symmetric
spaces based on the $G_2$ group manifold.
The anti-invariant set of involutive automorphism $\theta$ is $p
={{E-\theta E\theta}\o 2}$ and is an eight-parametric space. The form of $p$
is given by
\be
p=\pmatrix{\rho _1&\eta _1&\eta _4&0&\eta _5&\s \eta _3&\eta _2\cr
-\eta _1&-\rho _2&0&\eta _4&-\eta _6&-\s \eta _5&-\eta _3\cr
-\eta _4&0&\rho _2&\eta _1&-\eta _3&-\s \eta _5&-\eta _6\cr
0&-\eta _4&-\eta _1&-\rho _1&\eta _2&\s \eta _3&\eta _5\cr
\eta _5&\eta _6&\eta _3&\eta _2&\rho _1+\rho _2&\s \eta _1&0\cr
-\s \eta _3&-\s \eta _5&-\s \eta _5&-\s \eta _3&-\s \eta _1&0&\s \eta _1\cr
\eta _2&\eta _3&\eta _6&\eta _5&0&-\s \eta _1&-\rho _1-\rho _2
},
\ee
where
$$
\rho _1=2(h_1+h_2),\rho _2=2(h_1-h_2),$$\be
\eta _1={{e_1-\t e_1} \o 2},\eta _2={{e_2+\t e_2} \o 2},
\eta _3={{e_3-\t e_3} \o 2},\eta _4={{e_4-\t e_4} \o 2},
\eta _5={{-e_5-\t e_5} \o 2},\eta _6={{e_6-\t e_6} \o 2}.
\ee
Moreover, we have $[p,p]\subset f$ and $[f,p] \subset p$.
These properties of $\theta$, in the context of dual Riemannian symmetric spaces
is enough for construction of the unique pair of dual spaces. In fact,
the $\theta$-invariant algebra $f$ is the maximal compact subalgebra of $g_2^*$
and gives rise to the compact-noncompact dual Riemannian symmetric pairs
${{G_2} \o F}$ and
${{G_2^*} \o F}$.
\vskip 0.5cm
\noindent{\bf{Other Exceptional Gauged WZW Models}}

In this section, we give the involutive automorphisms $\theta$ which
according to equations (\ref{THETA}), convert the $g$ and $\bar A$ fields of
the vector gauged $F_4,E_{6,7,8}$ \w models to the corresponding chiral models,
\bea
\theta_{F_{4}}&=&\sum _{i=1}^{26}(e_{i,i+26}-e_{i+26,i}),
\nn\\
\theta_{E_{6}}&=&\sum _{i=1}^{39}(e_{i,79-i}-e_{i+39,40-i}),
\nn\\
\theta_{E_{7}}&=&\sum _{i=1}^{67}e_{i,68-i}-\sum _{i=1}^{66}e_{i+67,134-i},
\nn\\
\theta_{E_{8}}&=&\sum _{i=1}^{124}(e_{i,i+124}-e_{i+124,i}),
\eea
where $e_{ij}$ is the matrix with one at the $(i,j)$ entry and zero elsewhere.
It can easily be seen that the above automorphisms satisfy $\theta ^2=1$.
Moreover, the set of fixed points of the above involutive automorphisms is,
$f_{F_{4}}=sp(3)\otimes su(2),\,f_{E_{6}}=su(6)\otimes su(2),\,
f_{E_{7}}=su(8),\,f_{E_{8}}=so(16)$ which
shows that $\theta _{F_{4}},\theta _{E_{6}},\theta _{E_{7}},\theta _{E_{8}}$
belongs to the $F_4,E_6,E_7,E_8$ Lie group respectively.
\vskip 0.5cm
\noindent{\bf{Concluding Remarks}}

The vector gauged \w theory with any exceptional group $G$ gauged with any of
its null subgroups is
equivalent to the chiral gauged \w theory. The equivalence map is the
inner
automorphism of algebra ${\cal L}(G)$ which gives the irreducible Riemannian
globally symmetric spaces.
For example, in the case of $G_2$ group, the equivalence map gives the
irreducible Riemannian globally symmetric spaces  ${{G_2}\o {SU(2)\otimes
SU(2)}}$ and ${{G_2^*}\o {SU(2)\otimes SU(2)}}$.
Previously, the equivalence of the vector and chiral gauged \w theories with
any classical Lie groups were established. In all
the classical series, the corresponding map is an involutive automorphism of
the Lie algebra
\cite {4}.
\vskip 0.5 cm
\noindent{\bf{Appendix}}

We take the generators of $g_2$ Lie algebra as in \cite{8},
\bea
E_1&=&\rsi (e_{12}+e_{34})+\rth (e_{56}+e_{67}),\nn\\
E_2&=&\rtw (e_{17}+e_{54}),\nn\\
E_3&=&\rth (e_{16}-e_{64})+\rsi (e_{53}-e_{27}),\nn\\
E_4&=&\rtw (e_{13}+e_{24}),\nn\\
E_5&=&-\rsi (e_{15}+e_{74})+\rth (e_{26}+e_{63}),\nn\\
E_6&=&\rtw (e_{73}-e_{25}),
\eea
and $E_{-i}={\t E}_i$. The generators of Cartan subalgebra are given by
\bea
H_1&=&\cha (e_{11}-e_{22}+e_{33}-e_{44})+\rth (e_{55}-e_{77}),\nn\\
H_2&=&\frac{1}{4}(e_{11}+e_{22}-e_{33}-e_{44}).
\eea
We note the null property of generators as follows,
\be
E_1^3=E_2^2=E_3^2=E_4^2=E_5^3=E_6^2=
Tr (E_1^2)=Tr (E_5^2)=0.
\ee

\end{document}